\documentclass[
  reprint,
  amsmath,amssymb,
  longbibliography,
  aps, prl
]{revtex4-2}

\usepackage{dsfont}
\usepackage{graphicx}
\usepackage{overpic}
\usepackage{dcolumn}
\usepackage{bm}
\usepackage{xcolor}
\usepackage{epstopdf} % from supplement, if you still use eps

\usepackage[hidelinks]{hyperref}

\begin{document}

\preprint{}

\title{%
A dynamical order parameter for the transition to nonergodic dynamics in the discrete nonlinear Schrödinger equation
}

\author{Andrew Kalish}
\author{Pedro Fittipaldi de Castro}
\author{Wladimir A. Benalcazar}
\email{benalcazar@emory.edu}
\affiliation{Department of Physics, Emory University, Atlanta, Georgia 30322, USA}
\date{\today}

\begin{abstract}
The discrete nonlinear Schrödinger equation (DNLSE) exhibits a transition from ergodic, delocalized dynamics to a weakly nonergodic regime characterized by breather formation; yet, a precise characterization of this transition has remained elusive. By sampling many microcanonically equivalent initial conditions, we identify the asymptotic ensemble variance of the Kolmogorov–Sinai entropy as a dynamical order parameter that vanishes in the ergodic phase and becomes finite once ergodicity is broken. The relaxation time governing the ensemble convergence of the KS entropy displays an essential singularity at the transition, yielding a sharp boundary between the two dynamical regimes. This framework provides a trajectory-independent method for detecting ergodicity breaking that is broadly applicable to nonlinear lattice systems with conserved quantities.

\end{abstract}

\maketitle

\date{\today}

The one-dimensional discrete nonlinear Schrödinger equation (DNLSE) exhibits two thermodynamically distinct regions separated by an infinite-temperature line~\cite{Gradenigo_2021_partitionfunction}: a positive-temperature Gibbs phase, where canonical and microcanonical ensembles are equivalent, and a non-Gibbs region where the canonical partition function diverges and only the microcanonical ensemble is well defined.

Throughout the Gibbs phase, the long-time dynamics are delocalized and appear ergodic. Within the non-Gibbs region, approximate microcanonical entropy calculations show that energy localization becomes favored at high energies, concentrating a finite fraction of the total energy onto one or a few sites~\cite{Rasmussen_2001, Gradenigo_2021_partitionfunction}. At sufficiently large energy densities, these localized configurations (breathers) become dynamically stable, leading to weakly nonergodic behavior, in which microcanonically equivalent initial conditions fail to explore the same regions of phase space~\cite{Nonergodic_2018}. Despite these advances, a precise characterization of ergodicity breaking in the DNLSE as a dynamical phase transition remains lacking.

In this Letter, we approach the problem from an ensemble perspective. Instead of diagnosing ergodicity breaking from the statistics of individual trajectories, we sample many initial conditions at fixed energy and norm and monitor how their finite-time Kolmogorov–Sinai (KS) entropies evolve. This reveals how dynamical heterogeneity develops across the microcanonical manifold and allows us to sharply distinguish ergodic from nonergodic behavior. We show that ensemble variance of the KS entropy acquires a finite asymptotic value only when energy becomes trapped in sufficiently stable breather excitations, and that the corresponding relaxation time diverges with an essential singularity at the transition. This construction yields a trajectory-independent dynamical order parameter and provides a general tool for identifying ergodicity-breaking transitions in nonlinear Hamiltonian lattices.

The DNLSE, interchangeably referred to as the discrete Gross-Pitaevskii equation, or semiclassical Bose-Hubbard model, is a semiclassical mean-field model describing weakly interacting bosons in discrete geometries, such as Bose-Einstein condensates confined to optical lattices. Its equation of motion is
\begin{equation}
    i\dot{\psi}_j = -h(\psi_{j+1} + \psi_{j-1}) - g|\psi_j|^2 \psi_j, \,\,j=1, ..., N
    \label{eq:eom}
\end{equation}
with $g, h > 0$, describing the time evolution of a complex-valued field $\psi=(\psi_1, ..., \psi_N)\in\mathbb{C}^N$ on an $N$-site 1-dimensional lattice with periodic boundary conditions, where squared amplitudes $|\psi_j|^2$ represent the particle density on the $j$th lattice site. The DNLSE has broad relevance in physics beyond Bose condensates; notably, it is also utilized as a model of light propagation in waveguide arrays with Kerr nonlinearity \cite{optics, Fu_2022}, and has found diverse applications elsewhere, including plasma physics \cite{Plasma}, acoustics \cite{acoustics}, and biophysics \cite{DNLSEBiophysics}.

\begin{figure}
    \centering
    \includegraphics[width=\linewidth]{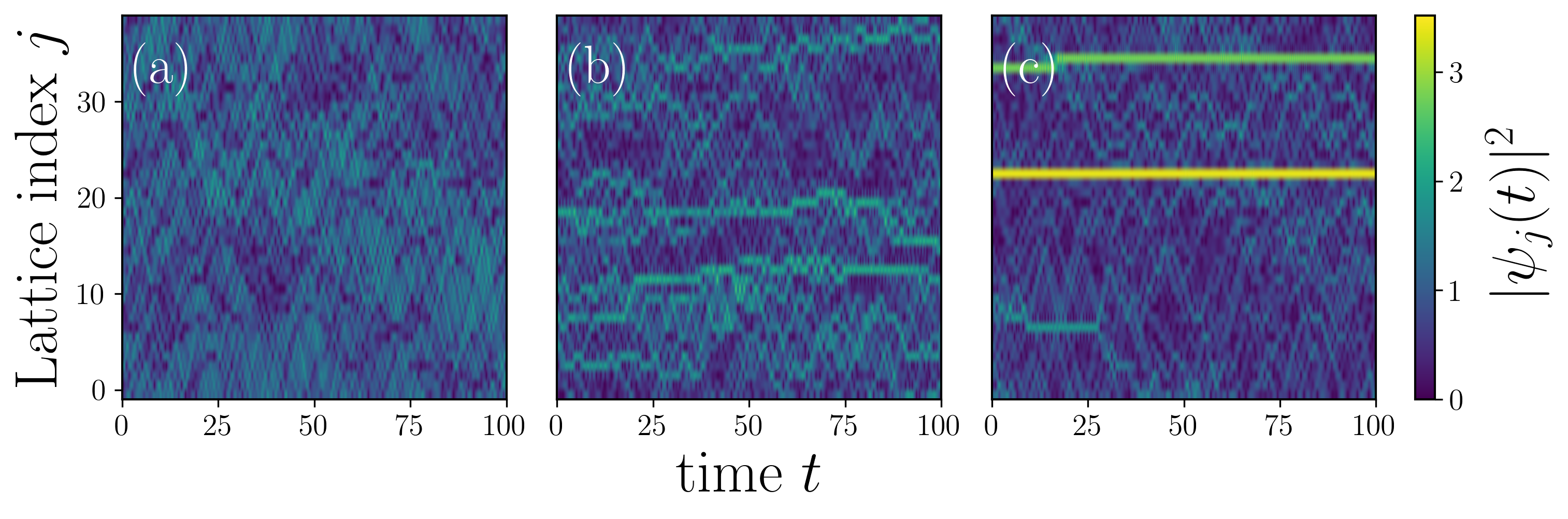}\\
    \includegraphics[width=\linewidth]{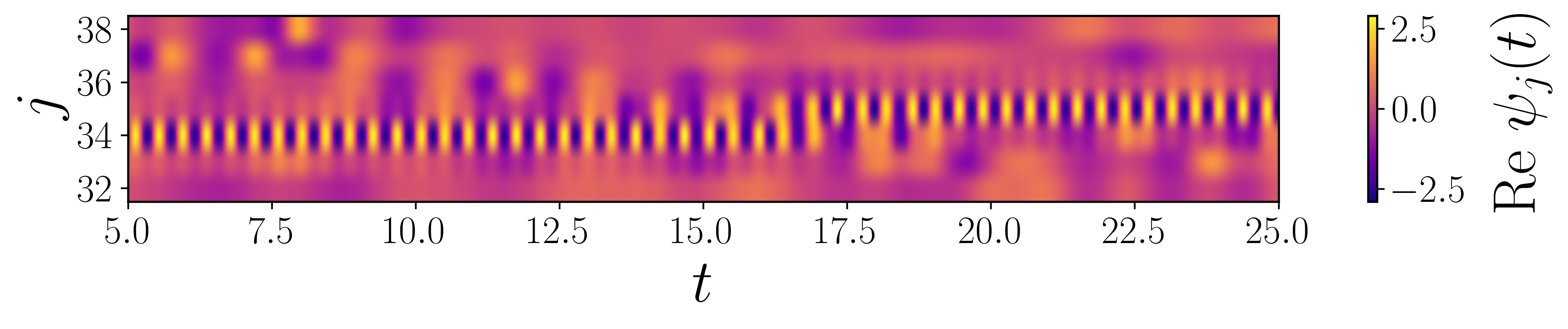}
        \caption{\textbf{Time evolution of a state $\psi$ under the discrete nonlinear Schrödinger equation}, Eq.~\eqref{eq:eom}, shown for (a) delocalized, (b) pseudo-localized, and (c) localized phases. The lattice size $N=40$ and amplitude density $a=1$ are fixed while energy densities are $e = 0.8, \,2.3, \,5.5$ from left to right. Bottom: Real component $\mathrm{Re} \,\psi_j(t)$ of the smaller breather evolution in $c$.}
    \label{fig:time_evolutions}
\end{figure} 

The system of equations \eqref{eq:eom} is invariant under global $U(1)$ phase rotation as well as time translation. The former symmetry corresponds with conservation of total particle number $||\psi||^2 = \sum_{j=1}^N|\psi_j|^2$, while the latter implies conservation of the Hamiltonian
\begin{equation}
    H = \sum_{j=1}^N h(\psi_j^*\psi_{j+1} +\psi_{j+1}^*\psi_j) + \frac{g}{2}|\psi_j|^4.
    \label{eq:ham}
\end{equation}
Here $\{\psi_j, i\psi_j^*\}$ are the conjugate variables and $i\dot{\psi_j} =-\partial H / \partial \psi_j^*$ recovers the equation of motion \eqref{eq:eom}. The thermodynamic state is specified by the conserved densities $a = ||\psi||^2/N$ and $e = H/N$. Throughout this work, we set $h=1$ and $g=2$, and focus on the $a=1$ slice of the phase diagram shown in Fig.~\ref{fig:phasediag}. 
\begin{figure}
\includegraphics[width=.49\linewidth]{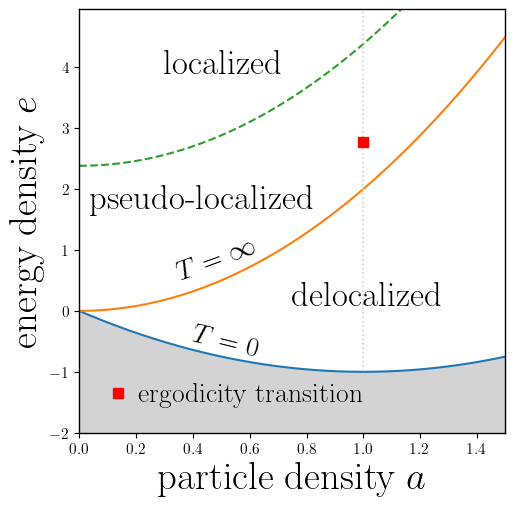}\includegraphics[width=.49\linewidth]{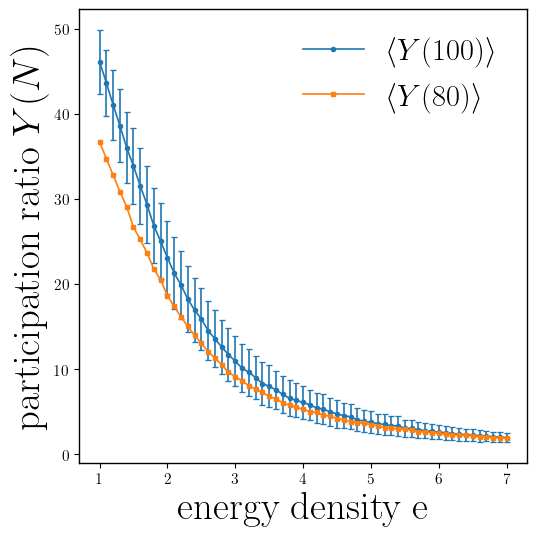}
\caption{\textbf{Phase diagram in $(e, a)$ space}. Ground state and infinite temperature isotherm was identified in \cite{Rasmussen_2001}. Breather localizations are still persistent in the pseudo-localized phase; however, the participation ratio (right, shown for $a=1$) is extensive for all states below the green dashed curve identified by \cite{Gradenigo_2021_partitionfunction}, which approaches the $T=\infty$ boundary as $N\rightarrow \infty$. Evolutions in each phase are shown in Fig.~\ref{fig:time_evolutions}. Left: Ensemble averages and standard deviations of the participation ratio $Y(N)$}
\label{fig:phasediag}
\end{figure}

Given the pair of conserved quantities in the DNLSE, we seek to determine for which values of $(e, a)$ is the microcanonical set $\mathcal{M}^{(N)}_{e, a} := \{\psi \in \mathbb{C}^N : H(\psi)/N= e, ||\psi||^2/N = a\}$ ergodic. For ergodic states, a relaxation time can be quantified with respect to the rate at which the variance of some observable across all states in $\mathcal M$ vanishes, and for nonergodic states there must exist some observable whose distribution maintains a finite variance asymptotically. 

Prior quantitative studies of ergodicity breaking in the DNLSE have relied on single-trajectory statistics of excursion times off of Poincaré manifolds \cite{Nonergodic_2018}, which, while demonstrating nonergodic behavior, face significant limitations. Our complementary approach - involving global statistics across hundreds of microcanonically equivalent initial conditions - offers a clearer picture of the transition. In particular, our framework allows for characterizing the divergence in relaxation times as the transition is approached, and further provides a metric of the "strength" of broken ergodicity throughout the nonergodic phase. 

Our approach to the study of ergodicity breaking in this system begins with sampling initial conditions from the microcanonical ensemble; that is, assigning a uniform probability weight to all states in $\mathcal{M}^{(N)}_{e, a}$. For even small values of $N$ an exact sampling of $\mathcal M$ is not feasible, and a Markov chain Monte-Carlo method is appropriate, which we discuss first. For convenience, we define the function $F:\mathbb{C}^N \rightarrow \mathbb{R}^2$, $F(\psi) := (H(\psi), ||\psi||^2) = (E, A)$, which maps a state to its conserved quantities. (We may sometimes abuse notation and also write $A$ to mean the function $A(\psi) = \|\psi||^2$, however, this should be clear from context.) Note that the microcanonical partition function may be written as 
\begin{equation}
\label{eq:coarea}
    \int_{\mathbb{\mathbb{C}}^{N}} \delta\!\big(F(\psi)-(E,A)\big)\,d\psi
=\int_{F^{-1}(E,A)} \frac{d\mathcal{\sigma}(\psi)}{\|\nabla H(\psi)\wedge\nabla A(\psi)\|},
\end{equation}
where $d\psi = \prod_jd(\rm Re \psi_j)\cdot d\rm (Im \psi_j)$, $d\sigma$ is formally the $2N-2$ dimensional Hausdorff measure, and $F^{-1}(E, A) =\mathcal{M}_{e, a}^{(N)}$ is a differentiable manifold at almost every $(E, A)$ by Sard's theorem \cite{Federer, coarea, Sard1942}. To sample from the microcanonical ensemble, we implement a random walk on the manifold $\mathcal{M}_{e, a}^{(N)}$ with an acceptance probability of a step from $\psi\rightarrow \psi'$ being 
\begin{equation}
    p(\psi\rightarrow\psi') =\min\left(1, \frac{\|\nabla H(\psi')\wedge\nabla A(\psi')\|}{\|\nabla H(\psi)\wedge\nabla A(\psi)\|}\cdot\frac{\alpha(\psi'\rightarrow\psi)}{\alpha(\psi\rightarrow\psi')}\right),
\end{equation}
where the factors $\alpha$ correct for any asymmetry in the proposals. Put concisely, we implement a Metropolis-Hastings chain \cite{metropolis1953, hastings1970} targeting the stationary distribution $p(\psi)\propto 1/||\nabla H(\psi)\wedge \nabla A(\psi)||$. Our method of implementing this sampling procedure, in particular our approximation of $\alpha$, and further details about this representation of the partition function are included in the supplementary material. Sampling in this manner gives a finite set of microcanonically weighted states which serve as a representative subset of $\mathcal M$. 

For a range of energies, we sample between many initial conditions from $\mathcal{M}_{e, a}^{(N)}$, and for each sampled state we compute the finite-time Lyapunov spectrum $\{\lambda_j(t)\}_{j=1}^{2N}$ and the Kolmogorov-Sinai entropy $S_{KS} \equiv \sum_{\lambda_j > 0} \lambda_j(t)$. 

Lyapunov exponents are real-valued with units of inverse time, and describe local stretching rates about a trajectory - with each $\lambda_j$ corresponding to an orthogonal degree of freedom in the tangent space of $\psi(t)$. They can be thought of as factors describing the strength of chaos, where commonly a single positive Lyapunov exponent is considered a sufficient condition for dynamics to be deemed chaotic.  

Zero-valued Lyapunov exponents correspond to conserved quantities, and near-zero exponents appear in the presence of "slow" degrees of freedom. The presence of breather localizations alters Lyapunov exponents in several ways, primarily by lowering energy of the background. This is similar in spirit to Rumpf's argument \cite{RumpfSimple} for how the formation of a breather modifies the thermodynamic entropy. A less obvious but important consequence of breathers on the Lyapunov spectra was found by Iubini and Politi \cite{Iubini_2021_chaos}, who showed that a region between a pair of breathers will locally have nearly conserved quantities, increasing the number of near-zero exponents. 

The sensitivity of the Lyapunov spectrum to the number, size, and position of breathers makes it a well-suited observable for quantifying the dynamical heterogeneity in the DNLSE. Studying the time-dependent variance of Kolmogorov-Sinai entropy $S_{KS} = \sum_{\lambda_j > 0} \lambda_j$ across ensemble samples therefore concisely quantifies this heterogeneity from a global perspective. Kolmogorov-Sinai entropy measurements and representative Lyapunov spectra are shown in Fig.~\ref{fig:lyps}. 

We utilize the algorithm due to Benettin \emph{et al.} and Shimada \emph{et al.} \cite{Benettin, Shimada} to compute the Lyapunov spectra. Details of our implementation are included in the supplementary material. For a comprehensive modern overview of the theory and methods of Lyapunov spectra, we refer to Ref. \onlinecite{Ginelli_2013}.

\begin{figure}
    \centering
    \includegraphics[width=.8\linewidth]{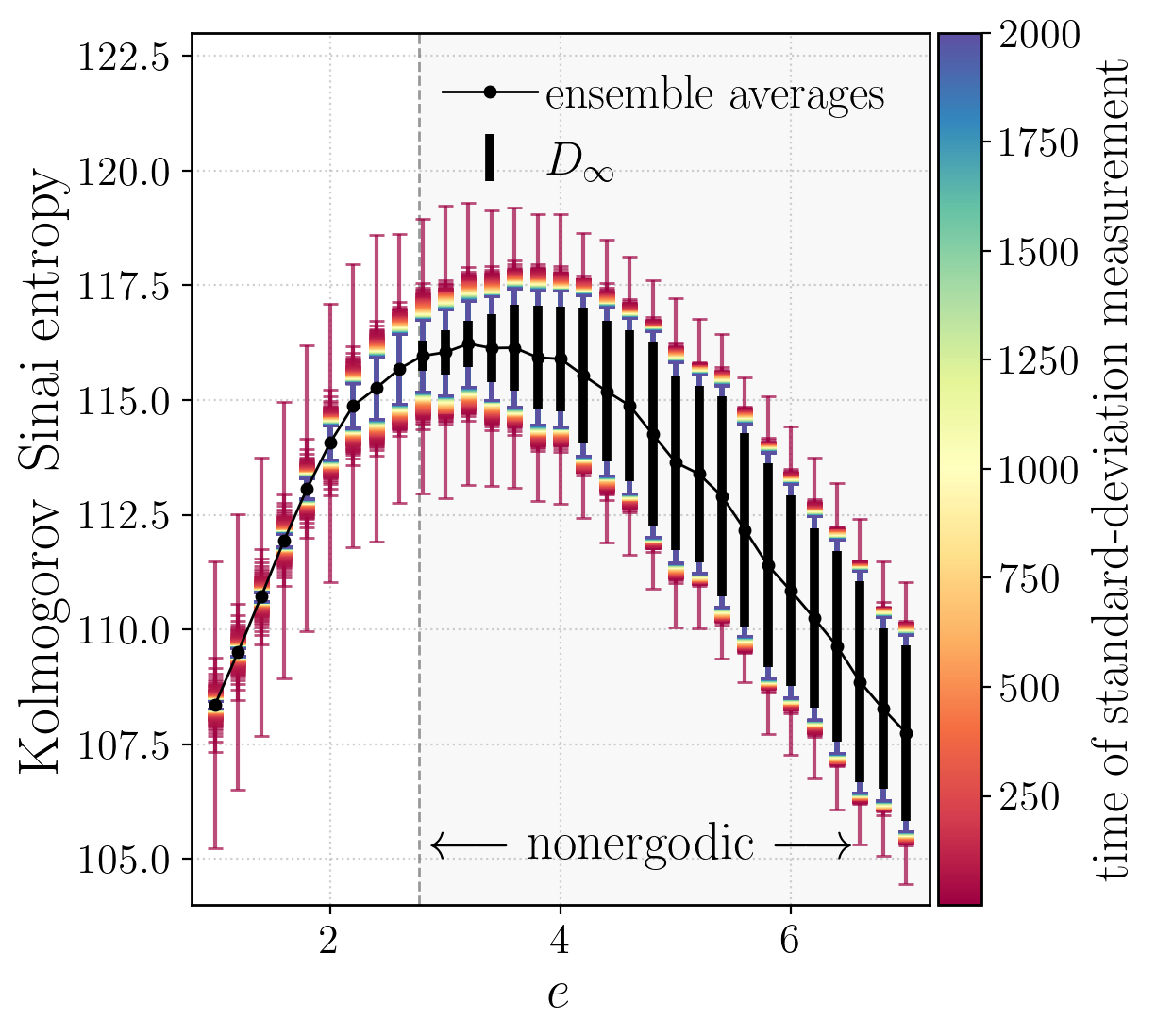}\\
    \centering
        \includegraphics[width=0.48\linewidth]{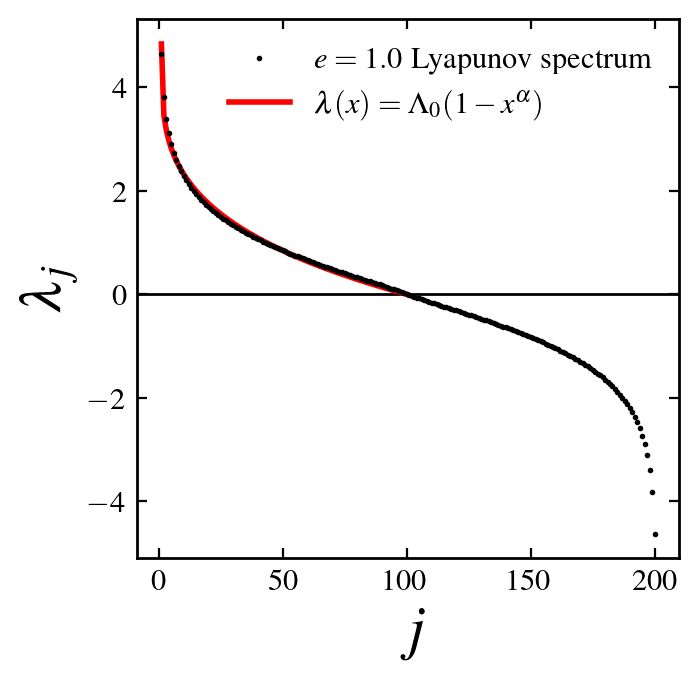}\includegraphics[width=0.48\linewidth]{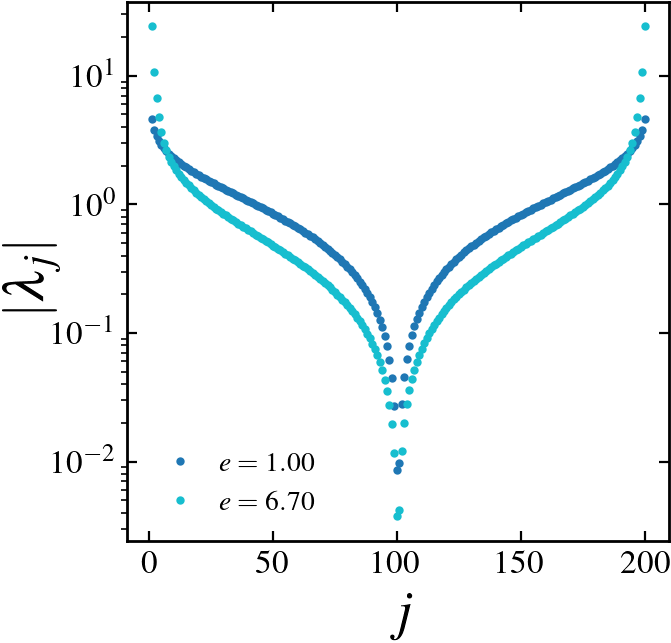}
    \caption{\textbf{Ensemble Lyapunov spectra, Kolmogorov-Sinai entropy and standard deviation measurements}. Top: ensemble sampled Kolmogorov-Sinai entropy and standard deviation measurements across the $a=1$ slice of the phase diagram. Vertical black bars denote predictions of $D_\infty$. Bottom: sample Lyapunov spectra. Left image shows the $e=1$ spectrum and the two-parameter fit found by \cite{Iubini_2021_chaos}. Right image shows the magnitudes of the same $e=1$ spectrum contrasted with the $e=6.7$ spectrum on a log scale. }
    \label{fig:lyps}
\end{figure}

At a point $\psi(t_0)$ on the manifold $\mathcal M$, there exists a $2N-2$ dimensional tangent space $T_{\psi(t_0)}\mathcal{M}$. The task of a Lyapunov–spectrum measurement is to identify a decomposition of the tangent space into subspaces
\begin{equation}
T_{\psi(t_0)} \mathcal{M} \cong \mathcal{O}_1 \oplus \cdots \oplus \mathcal{O}_m,
\end{equation}
where $m\leq 2N-2$, and for any $v\in \mathcal{O}_j$ and $t_0$ chosen arbitrarily, we expect an infinitesimal perturbation $\overline{\psi}(t_0) = \psi(t_0) + \varepsilon v$, ($\varepsilon>0$) to diverge from $\psi(t_0)$ at a rate 
\begin{equation}
d(\psi(t), \overline{\psi}(t)) \approx e^{\lambda_j (t-t_0)}d(\psi(t_0), \overline{\psi}(t_0)),
    \label{eq:exponentialdivergence}
\end{equation} 
where the metric between states here is \begin{equation}
    d(\psi(t), \overline{\psi}(t)) \equiv \frac{1}{2}\langle \psi(t) - \overline{\psi}(t)| \psi(t) - \overline{\psi}(t)\rangle
\end{equation}
\cite{PhysRevA.83.043611} and with $\lambda_j$ the same for any such $v$. $\lambda_j$ is called the characteristic Lyapunov exponent corresponding to the Oseledets subspace $\mathcal{O}_j$, and the dimension of $\mathcal{O}_j$ corresponds to the multiplicity of $\lambda_j$. For initial conditions with additional symmetries (i.e., a plane-wave) the Lyapunov exponents can appear in degenerate rows corresponding to larger dimensional Oseledets spaces, however for generic initial conditions $\mathcal O_j$ are typically one-dimensional. 

The procedure for determining this decomposition and the corresponding Lyapunov spectrum involves an iterative process along a trajectory which frequently measures stretching factors in the immediate neighborhood of the trajectory, and yields finite-time Lyapunov exponents $\{\lambda_j(t)\}$ which together account for the chaos accumulated along the trajectory up to time $t$. The multiplicative ergodic theorem \cite{Oseledets_original, Oseledets_notes} guarantees the convergence $\{\lambda_j(t)\} \rightarrow \{\lambda_j\}$ to values invariant of initial conditions when the dynamics are ergodic. 

\begin{figure}
    \centering
    \includegraphics[width=.49\linewidth]{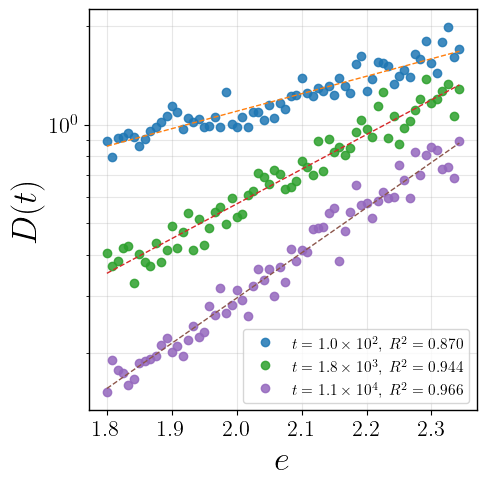}\includegraphics[width=.49\linewidth]{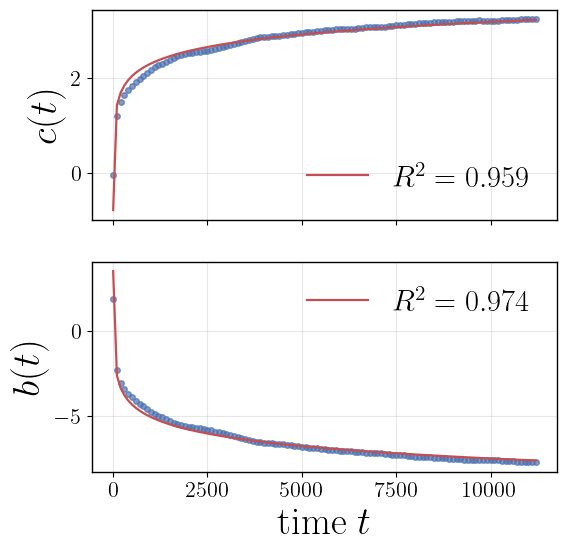}
    \includegraphics[width=.49\linewidth]{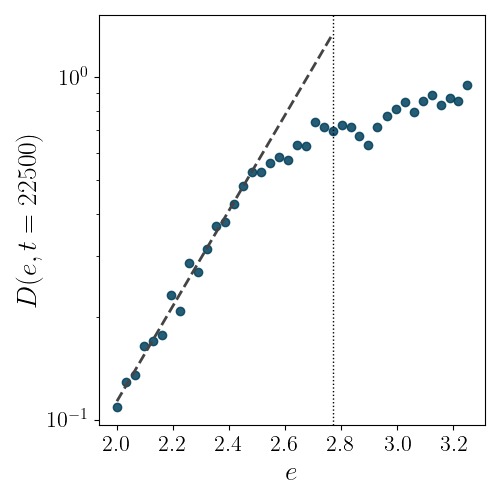}\includegraphics[width=.49\linewidth]{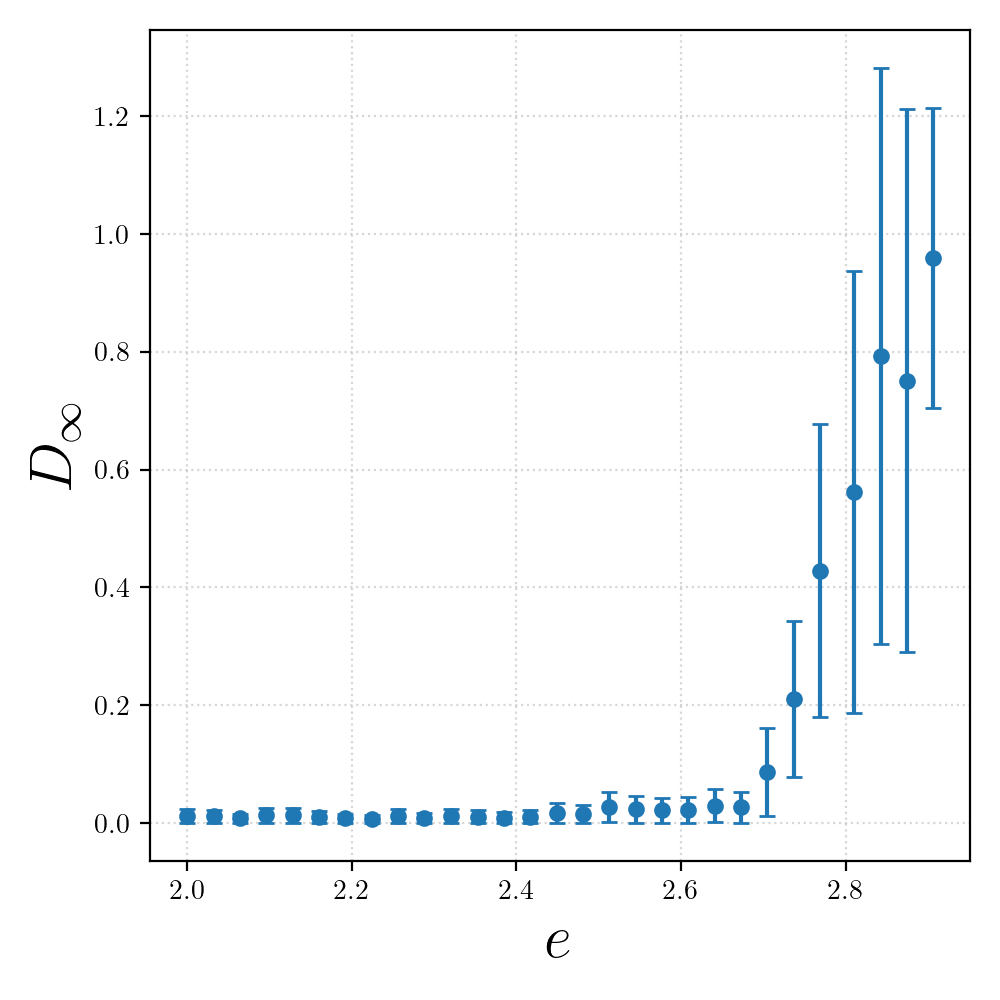}
    \caption{\textbf{Finite-time ensemble variance and infinite time predictions of Kolmogorov-Sinai entropy vs energy density.} Top left: standard deviation of the sum of finite time positive Lyapunov exponents across $150$ ensemble-sampled trajectories, after averaging over $100$, $1800$, and $11200$ time-units respectively, with $N=100$ and $a=1$. The fits are exponential curves $D(t)=\exp[b + ce]$, where $b=b(t)$, $c=c(t)$ both satisfy logarithmic growth laws shown on the top right. We observe that $c(t)=c_0+C\log(t)$, $b(t) = b_0-B\log(t)$, with $c_0\approx-0.32$, $b_0\approx2.24$, $C\approx0.38$, $B\approx1.07$.  Bottom left: breakdown in functional form of $D(e, t)$ near $e=2.5$. Bottom right: mean and standard deviation of posterior probability estimates of $D_\infty$.}
    \label{fig:finite_time_KS}
\end{figure}

We determine the statistics of the Lyapunov spectra for a wide range of energy densities spanning the $a=1$ slice of the phase diagram at $N=100$. Denoting the standard deviation of the finite-time Kolmogorov-Sinai entropy at time $t$ for states of energy $e$ as 
\begin{equation}D(e, t) \equiv \rm std_{ \{ \psi\in\mathcal{M}_{e, a=1}^{(N=100)}
\}} [S_{KS}(\psi)]. \label{eq:DDef} \end{equation} 
A sufficient condition for broken ergodicity is $\lim_{t\rightarrow \infty} D(e, t) > 0$; since if this limit were nonzero and the dynamics were ergodic it would contradict Oseledets multiplicative ergodic theorem. Furthermore, when the limit is finite its value serves as a metric of the severity of the violation of ergodicity. 

We observe across all energies that the standard deviation of Kolmogorov-Sinai entropy is well described by a mixed algebraic and exponential decay law of the form 
\begin{equation}
    D(e, t) = D_\infty +K_1\cdot t^{\gamma} + K_2 \exp(-t/\tau_{\rm exp}),
\end{equation}
where the parameters vary with energy. We fit $D(e, t)$ to this form and obtain posterior distributions for each parameter given the finite-time data using a Hamiltonian Monte Carlo algorithm \cite{HMC}. For all $e < 2.7$, the posterior mass of $D_\infty$ is strongly concentrated about $0$, consistent with the expectation of ergodicity in this region. Near $2.7$, the distributions broaden but remain within error of zero; and by $e = 2.75$ ergodicity is broken with near-certainty as the posterior density of $D_\infty = 0$ becomes negligible. Posterior means and standard deviations of $D_\infty$ in this window are shown in Fig~\ref{fig:finite_time_KS}.  

We find that for all $e < 2.5$, $K_2$ is negligible, and $\gamma$ increases linearly with energy, such that 
\begin{equation}
    D(e, t) = K_1\cdot t^{e\cdot C- B}, \,\,e \lesssim 2.5.
    \label{eq:D(t)}
\end{equation}
This relation was established first from the observation of an exponential dependence in energy at fixed times, as shown in the top left of Fig.~\ref{fig:finite_time_KS}, giving the form $D(e) = \exp(ec +b)$. Repeating these fits across time, we find that $c=c(t)$ and $b = b(t)$ satisfy logarithmic laws of the form $c(t) = c_0 + C\log(t)$ and $b(t) = b_0 - B\log(t)$, shown in the top right of Fig.~\ref{fig:finite_time_KS}. Altogether, these relations reduce to Eq.\eqref{eq:D(t)}, with $K_1 = \exp(ec_0 + b_0)$. The constants $C$ and $B$ suggest a critical energy $e^* = B/C \approx2.77$ as the value at which $\gamma\rightarrow0$.

Importantly, $e^*$ is outside of the window where the purely algebraic functional form is satisfied by $D$. In the region $e\in[2.5,2.77]$, both $K_1$ and $K_2$ are nonzero, and the exponential term has a long characteristic timescale $\tau_{\rm exp} \sim 7\cdot10^4$ time units, but does not seem to grow significantly with energy. 

While the time necessary for $D\rightarrow 0$ exactly is generally infinite, one may define an effective relaxation time $\tau^{(\delta)}$ for some $\delta > 0$ as the smallest time such that $D(e, t) < \delta$ for all $t \geq \tau^{(\delta)}$. Since the term $K_2\exp(t/\tau_{\rm exp})$ has a long but not diverging characteristic timescale $\tau_{\rm exp}$, for small enough $\delta$ it is effectively negligible for the sake of determining $\tau^{(\delta)}$ and we may from Eq.\eqref{eq:D(t)} derive
\begin{equation}
    \tau^{(\delta)}(e) = \exp\left(\frac{(ec_0 + b_0-log(\delta))/C}{e^*-e}\right) 
    \label{eq:timediverging}
\end{equation}
suggesting an essential singularity in the relaxation time at $e^*$.

Timescales of decay in the region $e\in[2.5, 2.77]$ are extremely long, and competition between the exponential decay and algebraic decay in this region makes a precise determination of $\gamma$ from the 5-parameter fit alone difficult, where distributions of the parameters are generally multimodal. It is important to clarify that Eq. \eqref{eq:timediverging}, while confirmed in the region $e<2.5$, is an extrapolation assuming that the algebraic term holds that exact form in the critical window $e\in[2.5, 2.77]$; however, to prove this will require further investigation. 

Even if $\gamma$ cannot be cleanly resolved as of yet for $e\in [2.5, 2.77]$, the fact that $D_\infty$ is predicted to become nonzero at an energy within error of $e^*$ gives us confidence that the extrapolated linear vanishing of $\gamma$ is correct, and that the divergence in relaxation times at $e^*$ is driven by the extrapolated behavior of the algebraic decay. 

In conclusion, through a study of the convergence rates of Kolmogorov-Sinai entropy across a large set of microcanonical ensemble samples, our results significantly contribute to an understanding of the properties of the ergodicity transition in the DNLSE. In the sense that the transition from ergodic to nonergodic dynamics can be considered a phase transition, the asymptotic variance of Kolmogorov-Sinai entropy $\lim_{t\rightarrow\infty}D(e, t) = D_\infty(e)$ may be considered a dynamical order parameter; where $D_\infty = 0$ in the ergodic phase and $D_\infty > 0$ in the nonergodic phase. Additionally, it seems likely that $D_\infty$ emerges continuously from zero, so that in this sense the dynamical transition is continuous. 

While demonstrated for the DNLSE, the provided framework should apply quite generally to systems that support both ergodic and nonergodic behavior. Beyond the binary classification of a state as ergodic or not, this method could be useful in establishing universality classes of nonergodic systems, classified by the form of divergence in relaxation times.

\begin{acknowledgements}
PFC and WAB thank the support of startup funds from Emory University.
\end{acknowledgements}

\bibliography{refs}
\nocite{apsrev42Control}
\clearpage
\onecolumngrid

\begin{center}
\textbf{\large Supplemental Material for\\[2pt]
A dynamical order parameter for the transition to nonergodic dynamics in the discrete nonlinear Schrödinger equation}
\end{center}

% Optional: change numbering to have S1, S2, ... etc.
\setcounter{section}{0}
\setcounter{equation}{0}
\setcounter{figure}{0}
\setcounter{table}{0}
\renewcommand{\thesection}{S\arabic{section}}
\renewcommand{\theequation}{S\arabic{equation}}
\renewcommand{\thefigure}{S\arabic{figure}}
\renewcommand{\thetable}{S\arabic{table}}

\title{Supplemental Material for: 'A dynamical order parameter for the transition to nonergodic dynamics in the discrete nonlinear Schrödinger equation'}

\maketitle
Some background in the thermodynamics of the DNLSE is helpful for understanding the nature of the ergodic-nonergodic transition. 
\section{Thermodynamics of the DNLSE}
\label{sec:thermo}

Here we give a brief overview of the thermodynamics of the discrete nonlinear Schrödinger equation (DNLSE). Recall the Hamiltonian

\begin{equation}
 H[\psi] =
 \sum_{j=1}^{N}
 \left[
 h(\psi_j^*\psi_{j+1} + \psi_{j+1}^*\psi_j)
 +
 \frac{g}{2}|\psi_j|^4
 \right].
 \label{eq:H_supp}
\end{equation}
where the canonical variables are $\{\psi_j,i\psi_j^{*}\}$ and the resulting equation of motion is
\begin{equation}
 i\dot{\psi}_j = -h(\psi_{j+1} + \psi_{j-1}) - g|\psi_j|^2 \psi_j,
 \qquad j = 1,\dots,N.
 \label{eq:eom_supp}
\end{equation}

Total particle number
\begin{equation}
 A = \|\psi\|^2
 =
 \sum_{j=1}^{N}|\psi_j|^2,
\end{equation}
and the energy
$E = H[\psi]$ are conserved quantities. We frequently use the intensive energy and particle densities
\begin{equation}
 a = \frac{A}{N},\qquad
 e = \frac{E}{N}
\end{equation}

\subsection{Canonical and microcanonical ensembles}

In the grand-canonical ensemble, states are assigned probabilistic weights
\begin{equation}
 p_{\beta,\mu}(\psi) \propto
 \exp\big[-\beta(H(\psi)-\mu\|\psi\|^2)\big],
 \label{eq:grand_canonical}
\end{equation}
where $\beta$ and $\mu$ are Lagrange multipliers playing the role of
inverse temperature and chemical potential. In the thermodynamic limit,
Rasmussen \emph{et al.} \cite{Rasmussen_2001} identified both a zero-temperature ground state and an infinite-temperature isotherm, located at
\begin{equation}
 e_{\text{ground}}(a) = -2a + \frac{g}{2}a^2,\qquad
 e_{\infty}(a) = ga^2,
 \label{eq:isotherms}
\end{equation}
corresponding to $\beta=\infty$ and $\beta=0$, respectively. The
curve $e=e_\infty(a)$ is an infinite-temperature parabola.

The thermodynamic limit of the DNLSE exhibits a non-concave entropy
as a function of $(e,a)$, and for $\beta < 0$ the grand-canonical and
canonical ensembles become ill-defined because their partition functions diverge.
In this negative-temperature region the standard equivalence of
ensembles breaks down, and only suitably modified thermal ensembles or
the microcanonical ensemble remain well-defined.

In the microcanonical ensemble, states are uniformly weighted on the
intersection of fixed energy and fixed norm level sets,
\begin{equation}
 p_{e,a}(\psi) \propto
 \mathbf{1}_{\mathcal{M}_{e, a}^{(N)}}
 \label{eq:micro_def}
\end{equation}
where $\mathbf{1}_X$ is the indicator of the set $X$, and $\mathcal{M}_{e, a}^{(N)} = \{\psi\in\mathbb{C}^N : H(\psi)/N = e, \|\psi\|^2/N = a\}.$ For almost every $(e,a)$ this intersection forms a smooth
$(2N-2)$-dimensional manifold in $\mathbb{C}^N$.

\subsection{Explaining energy localization: the separable high-temperature model}
As shown in Fig.~2 (main text), the system becomes increasingly localized as energy density increases. Significant recent progress has been made towards an understanding of such phenomena from a statistical perspective, such as in \cite{Gradenigo_2021_partitionfunction, Gradenigo_2021_moredetailedpartitionfunction}, which we summarize here. 

At high energy densities (especially at and above the infinite
temperature parabola $e=ga^2$), the hopping part
\begin{equation}
 H_{\rm hop} =
 \sum_{j=1}^{N}
 (\psi_j^*\psi_{j+1} + \psi_{j+1}^*\psi_j)
\end{equation}
is subdominant in the microcanonical statistics. When the phases of
the complex amplitudes are uncorrelated, $H_{\rm hop}$ behaves approximately like a
sum of $N$ independent random variables with mean
$\langle H_{\rm hop} \rangle \approx 0$ and standard deviation
$\mathrm{std}(H_{\rm hop}) \propto \sqrt{N}$. In contrast, the nonlinear
term
\begin{equation}
 H_{\rm nl} = \frac{g}{2}\sum_{j=1}^N |\psi_j|^4
\end{equation}
is extensive at all finite energies, scaling as $O(N)$. The linear contribution to the total
energy is therefore generally subdominant and negligible at high energies compared to
the nonlinear contribution.

A standard approximation is thus to study the separable model with Hamiltonian $\widehat{H} = H_{\rm nl}$ subject to the same fixed norm constraint $\|\psi\|^2 = A$. The
thermodynamics of $\widehat{H}$ capture many qualitative features of
the full DNLSE in the high-energy and negative-temperature regime, but with analytically tractable statistics. Notably, the transition to negative temperatures occurs along the same isotherm \(e = ga^2\) for both $H$ and $\widehat{H}$. 

\subsection{Localization and participation ratio}

Useful for quantifying localization, the participation ratio of the
energy distribution is defined as
\begin{equation}
 Y(N) =
 \frac{\left(\sum_{j=1}^{N} \varepsilon_j \right)^2}
      {\sum_{j=1}^{N} \varepsilon_j^2}.
 \label{eq:PR_def}
\end{equation}
For a delocalized configuration, $Y(N)\propto N$, and in the fully localized limit where almost all the energy sits on one site, $Y(N)\rightarrow1$, independent of $N$. Note that this is the reciprocal of the quantity $Y_2(N)$ which in the literature is commonly also called the participation ratio. 

For the separable model $\widehat{H}$, Gradenigo \emph{et al.} solved
the microcanonical ensemble exactly. Writing the local energy as
\begin{equation}
 \varepsilon_j =
 \frac{g}{2}|\psi_j|^4,
\end{equation}
they showed that the single-site energy distribution has a stretched
exponential tail,
\begin{equation}
 P(\varepsilon_j) \propto
 \exp\big(-\mu\sqrt{\varepsilon_j}\big),
 \label{eq:stretched}
\end{equation}
with $\mu>0$ determined by the constraints $(e,a)$. For a sum of $N$
such terms constrained to a fixed total energy
$\sum_j \varepsilon_j = E$, at a sufficiently large energy densities there is a transition from roughly independent and identically distributed amplitudes to a symmetry-broken, localized phase with a single lattice site hosting macroscopic fractions of the system’s total energy. This transition is understood naturally with results of large-deviation theory applied to the fat-tailed marginal distribution of Eq. \eqref{eq:stretched}. 

Heuristically, the large-deviation theory argument can be understood as follows. One can see that since the tail of the distribution decays slowly, extremely large energy values are only slightly less probable than moderately large energies. So when forced to sum to a very large energy, configurations where a single site hosts a large amount of energy—such that the rest of the lattice can remain at a more probable low-energy state—become dominant. 

The results of Gradenigo \emph{et al.} predict that in the thermodynamic limit the localization transition coincides with the transition to negative temperatures at $e=ga^2$, while at finite size the full localization transition occurs at
\begin{equation}
 e \approx ga^2 + 11.05\,N^{-1/3},
 \label{eq:finite_size_threshold}
\end{equation}
with an intermediate pseudo-localized regime between
$e=ga^2$ and this finite-size threshold. 

This argument well explains the localization trend statistically. From a dynamical perspective, there are several intuitive justifications for breather stability. In order for energy to transfer from a breather to a neighboring site, there must be a phase resonance event where the phase of the breather locks with the phase of a neighboring site over a finite time interval. Since the frequency of oscillation in the complex plane at a site is proportional to the amplitude, breather frequencies are generally significantly greater than those of neighboring sites, making such resonance events rare, and accounting for stability of breathers only increasing with their magnitudes. A more quantitative understanding of dynamical stabillity is offered by principal component analysis of the dynamics of the breather and its neighboring sites, revealing a vanishing eigenvalue corresponding to a constrained subspace \cite{freezing} when the breather is placed on a positive temperature background (far from equilibrium). A mathematical study of the mean-field limit \cite{Arezzo_2022} gives further insight into this phenomena, using Morse theory to quantify this through a gap in the spectrum of the Laplace-Beltrami operator on the potential energy surface. 

Taken together, since large energy localizations in the form of rapidly oscillating breathers are statistically more likely as energy density increases, and the stability of such breathers increases with their energy, this picture intuitively justifies the breaking of ergodicity at high energies in the DNLSE. 

\subsection{Microcanonical geometry}
\label{sec:microcanonical}
Here we go over in some more detail the representation of the partition function which gives the proper sampling measure. Recall the microcanonical partition function 
\begin{equation}
 \Omega(E,A,N) = \int_{\mathbb{R}^{2N}}
 \delta\!\big(E - H(\psi)\big)\,
 \delta\!\big(A - \|\psi\|^2\big)\, d\psi,
 \label{eq:Z_micro_def}
\end{equation}
where $d\psi$ is the $2N$-dimensional Lebesgue measure and $\psi$ is considered as a $2N$-dimensional vector storing real and imaginary components at each site as distinct dimensions. Define $F:\mathbb{R}^{2N} \rightarrow \mathbb{R}^2$, 
\begin{equation}
 F(\psi) =
 \begin{pmatrix}
   H(\psi)\\
   \|\psi\|^2
 \end{pmatrix},
 \qquad
 F(\psi) = (E,A)
 \text{ on the constraint surface}.
\end{equation}

The coarea formula \cite{coarea, Federer} is a result of geometric measure theory which allows us to decompose the constrained integral in $\mathbb{R}^{2N}$ over the fibres (inverse images) of $F$,
\begin{equation}
 \int_{\mathbb{R}^{2N}} \varphi(\psi)\,J_F(\psi)\, d\psi
 =
 \int_{\mathbb{R}^2} \left(
 \int_{F^{-1}(E',A')} \varphi(\psi)\, d\sigma(\psi)\right)dE'\,dA'\,,
 \label{eq:coarea_with_J}
\end{equation}
valid for any measurable $\varphi:\mathbb{R}^{2N}\rightarrow\mathbb{R}$, where $d\sigma$ is the induced
$(2N-2)$-dimensional Hausdorff (surface) measure on the level sets
$F^{-1}(E',A')$, and 
\begin{equation}
 J_F(\psi) = \|\nabla H(\psi)\wedge\nabla A(\psi)\|
\end{equation}
is the Jacobian factor given by the norm of the two–form spanned by
the gradients of the conserved quantities.

We set
\begin{equation}
 \varphi(\psi)
 =
 \frac{\delta\!\big(F(\psi)-(E,A)\big)}{J_F(\psi)}
 =
 \frac{\delta\!\big(H(\psi)-E\big)\,
       \delta\!\big(\|\psi\|^2-A\big)}{J_F(\psi)}.
\end{equation}
so that the left-hand side of Eq.~\eqref{eq:coarea_with_J} is the microcanonical partition function in the form of \eqref{eq:Z_micro_def}, and the right-hand side becomes
\begin{equation}
 \int_{\mathbb{R}^2} dE'\,dA'\,
 \int_{F^{-1}(E',A')}
 \frac{\delta\!\big(F(\psi)-(E,A)\big)}{J_F(\psi)}\,
 d\sigma(\psi).
\end{equation}
For $\psi\in F^{-1}(E',A')$,  
\begin{equation}
 \delta\!\big(F(\psi)-(E,A)\big)
 =
 \delta(E'-E)\,\delta(A'-A)
\end{equation}
is constant on the level set. We can therefore
pull it out of the inner integral to obtain
\begin{equation}
 \int_{\mathbb{R}^2} dE'\,dA'\,
 \delta(E'-E)\,\delta(A'-A)
 \int_{F^{-1}(E',A')}
 \frac{d\sigma(\psi)}{J_F(\psi)}
 =
 \int_{F^{-1}(E,A)}
 \frac{d\sigma(\psi)}{J_F(\psi)},
\end{equation}
that is, 
\begin{equation}
 \Omega(E,A,N)
 =
 \int_{F^{-1}(E, A)}
 \frac{d\sigma(\psi)}{\|\nabla H(\psi)\wedge\nabla A(\psi)\|}.
 \label{eq:Z_surface_form_new}
\end{equation}

The same construction extends to $m$ mutually involutive conserved quantities
$\{V_k\}$ by replacing $F(\psi)=(H(\psi),\|\psi\|^2)$ with
$F(\psi)=(V_1(\psi),\dots,V_m(\psi))$ and
$\nabla H\wedge\nabla A$ with
$\nabla V_1\wedge\cdots\wedge\nabla V_m$. 

A proper microcanonical Metropolis-Hastings chain should therefore target a stationary distribution 
\begin{equation}
    p(\psi) = \frac{d\sigma(\psi)}{\|\nabla H(\psi)\wedge\nabla A(\psi)\|}.
    \label{eq:micro_measure}
\end{equation}
We first go over details of Lyapunov spectrum measurement, followed by an overview of the Metropolis-Hastings procedure. 

\section{Lyapunov spectrum statistics}
Here we discuss the technical details of our Lyapunov spectrum analysis. We begin with details of Lyapunov spectrum measurement, followed by the implementation of the Metropolis-Hastings chain from which initial conditions are sampled from. 

\subsection{Benettin, Shimada algorithm for Lyapunov spectrum computation}
We compute the full finite-time Lyapunov spectrum 
$\{\lambda_i\}_{i=1}^{2N}$ using the Benettin–Shimada algorithm 
\cite{Benettin,Shimada}. From an initial state $\psi(t_0)$ we construct 
$2N$ perturbed copies 
$\psi^{(i)}(t_0)=\psi(t_0)+\epsilon v^{(i)}$, where 
$\{v^{(i)}\}$ is an (initially arbitrary) orthonormal basis of $T_{\psi(t_0)}\mathcal{M}$. $\mathcal M$ abbreviates the microcanonical surface $F^{-1}(E, A)$. The tangent space has dimension $2N-2$, but keeping the full set of $2N$ vectors modifies the spectrum only by the addition of two zero Lyapunov exponents corresponding to the conserved quantities. 
These vectors $v^{(i)}$ form the columns of a matrix we call $Q_{(1)}$.  
We evolve $\psi(t)$ and all perturbations for a short interval 
$\tau_{GS}\!\sim\!0.07-0.1$ time-units, then subtract the reference trajectory to form the matrix  
\begin{equation}V_{(1)}=[v^{(1)}(t_0+\tau_{GS}),\ldots,v^{(2N)}(t_0+\tau_{GS})].\end{equation}  
A QR decomposition \begin{equation}
    V_{(1)}=Q_{(2)}R^{(1)}
\end{equation} is then performed: $Q_2$ is an orthonormal matrix whose columns are chosen as the next set of perturbations.
The diagonal elements of $R^{(1)}$ give local stretching factors, and the 
off-diagonal elements describe subspace mixing. Repeating this procedure along 
the trajectory yields sequences $\{Q_{(m)},R^{(m)}\}$, from which the finite-time 
exponents accumulate as
\begin{equation}
\lambda_i(t)=\frac{1}{t}\sum_{m=1}^{t/\tau_{GS}}
\frac{\ln R_{ii}^{(m)}}{\tau_{GS}} .
\label{eq:finite_time_lyap}
\end{equation}
Oseledets’ theorem ensures that the partial sums converge,  
$\lambda_j(t)\rightarrow \lambda_j$, for very general dynamical systems; and in an ergodic set $\lambda_j$ is independent of initial condition.

Two distinct timescales govern the convergence.  
The first, $\tau_A$, characterizes the alignment of perturbation vectors with 
the invariant subspaces as the off-diagonal elements of $R^{(m)}$ decay. This convergence occurs automatically as a consequence of the Gram-Schmidt algorithm used for the $QR$ decomposition to select the next orthonormal basis from the set of evolved perturbations. The orthogonalization begins by fixing the direction corresponding to the perturbation whose singular value in the decomposition is greatest, which will align closely with the most unstable direction regardless of the initial chosen basis. Iteratively, the directions corresponding to the smaller exponents are determined. 
\begin{figure}
    \centering
    \includegraphics[width=8.5 cm]{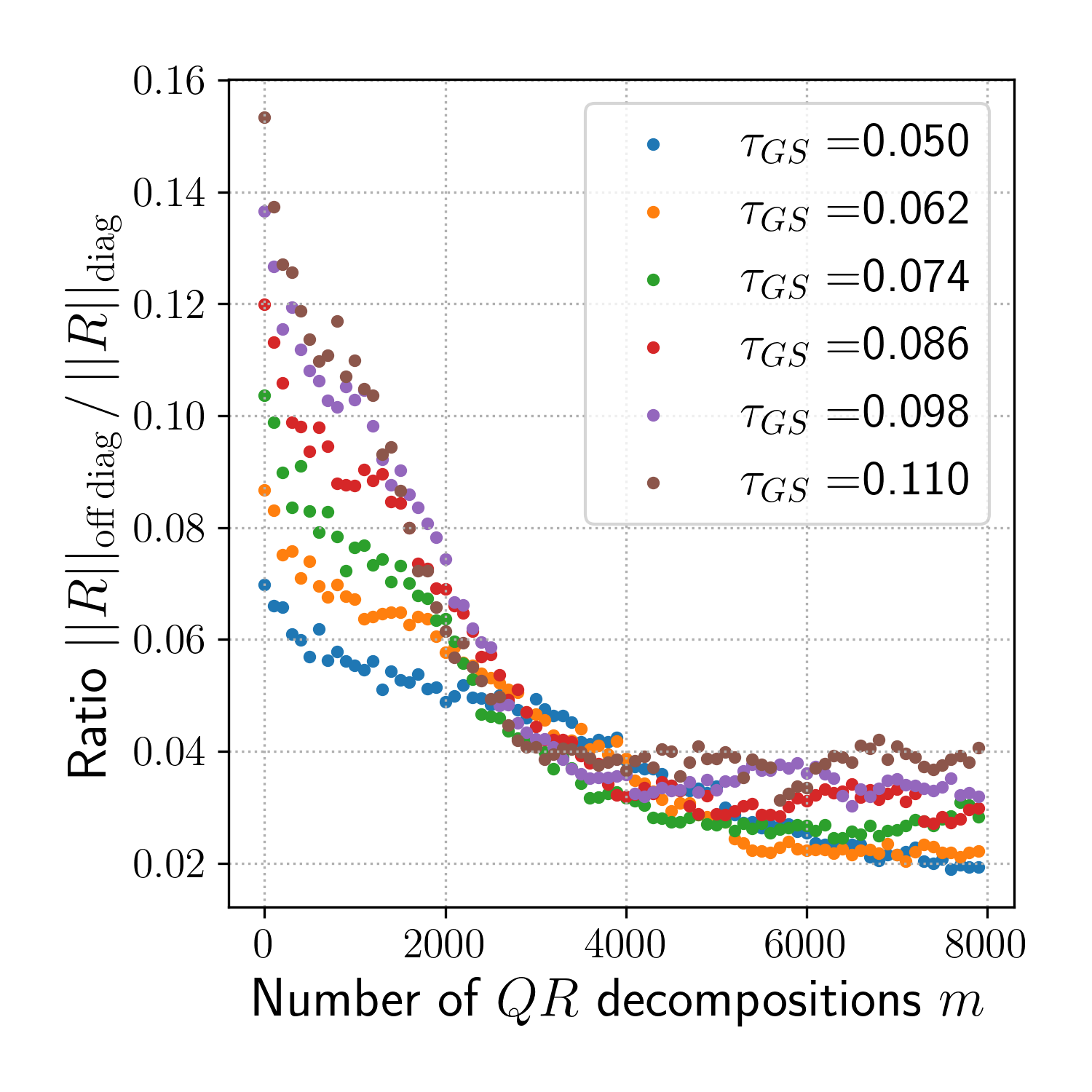}
    \caption{\textbf{Lyapunov spectrum convergence timescale $\tau_A$}. Ratio of matrix norms of off diagonal elements of the $R$ matrices compared to the diagonal elements falling to small values as Oseledets subspaces are iteratively identified, shrinking the off diagonal elements and containing chaotic stretching to the diagonal elements. For very small $\tau_{GS}$ one can achieve more optimal ratios, however a decreased $\tau_{GS}$ makes smaller integration timesteps of increased importance. We recommend setting $\tau_{GS} \geq 10\Delta t$ for an integration timestep $\Delta t$. The relaxation timescale $\tau_A$ is defined by the onset of
  the plateau.}
    \label{fig:placeholder}
\end{figure}
We estimate $\tau_A$ by tracking the ratio of diagonal to 
off-diagonal norms of $R^{(m)}$,
\begin{equation}
\begin{split}
\|R^{(m)}\|_{\text{diag}} &= \sum_i |R_{ii}|,\\
\|R^{(m)}\|_{\text{off}}  &= \sum_{i>j}|R_{ij}|,
\end{split}
\label{eq:Rstuff}
\end{equation}
which collapses towards small but nonzero values as the procedure iterates, as shown in Fig.\ref{fig:placeholder}. For short-time measurements – or more generally when one is particularly concerned about the instantaneous values of $\lambda_j(t)$ – one should only begin measuring the exponents after $t\geq \tau_A$. For very long measurements this is less of a concern. 

The second timescale, $\tau_B$, is the ergodic averaging time required for the finite-time exponents to converge to ensemble values; strictly speaking $\tau_B$ is almost necessarily infinite if one requires exact convergence, however one can discuss an effective time $\tau_B^{(\delta)}$ as the time where standard deviations across exponents are less than $\delta > 0$, as discussed in the text for Kolmogorov-Sinai entropy distributions. When dynamics are nonergodic there exists some $\delta > 0$ such that $\tau_B^{(\delta)}$ is not well defined. The exponents themselves will still converge to stationary values, however they will differ across initial conditions. Therefore ensemble averaging over many initial conditions as described below is therefore required to both obtain representative distributions and distinguish between the ergodic and nonergodic phase. 

\subsection{Microcanonical Metropolis-Hastings for sampling initial conditions}

To sample the manifold $\mathcal{M}_{e, a}^{(N)}$ (which we will denote $\mathcal{M}(E, A)$) according to the induced
measure \eqref{eq:micro_measure}, we implement a Metropolis–Hastings Markov chain with proposals constructed in the tangent space. Starting from a state $\psi\in\mathcal{M}(E,A)$, we draw a random vector
$\delta\psi$ orthogonal (in the real inner product) to both $\nabla H$
and $\nabla A$,
\begin{equation}
 \mathrm{Re}\big(\delta\psi^\dagger \nabla H\big) = 0,
 \qquad
 \mathrm{Re}\big(\delta\psi^\dagger \nabla A\big) = 0.
\end{equation}
For sufficiently small step size $\varepsilon$ the perturbed state
$\psi + \varepsilon\,\delta\psi$ stays close to $\mathcal{M}(E,A)$,
with deviations from the constraint surfaces of order
$O(\varepsilon^2)$.

We then project the trial state exactly back to the microcanonical manifold by a Newton iteration,
\begin{equation}
 \psi' = \mathrm{Proj}_{E,A}\big[\psi + \varepsilon\,\delta\psi\big],
 \label{eq:projection}
\end{equation}
using the two constraints $H(\psi')=E$ and $\|\psi'\|^2=A$. Because of the curvature of $\mathcal{M}(E,A)$, the proposal distribution is generally asymmetric: Denoting by $\alpha(\psi\rightarrow\psi')$ the probability
that a tangent step and projection from $\psi$ lands at $\psi'$, we generally have 
\begin{equation}
 \alpha(\psi'\rightarrow\psi) \neq \alpha(\psi\rightarrow\psi').
\end{equation}
The Metropolis–Hastings acceptance probability that yields the
stationary measure \eqref{eq:micro_measure} is
\begin{equation}
 p(\psi\to\psi')
 =
 \min\!\left\{
 1,\;
 \frac{J_F(\psi)}{J_F(\psi')}
 \frac{\alpha(\psi\rightarrow\psi')}{\alpha(\psi'\rightarrow\psi)}
 \right\},
 \label{eq:MH_accept}
\end{equation}
where the ratio of proposal probabilities is called the Hastings factor.

In practice, explicit evaluation of $\alpha(\psi\rightarrow\psi')/\alpha(\psi'\rightarrow\psi)$ is
costly. We approximate the Hastings factor by sampling small tangent
balls around $\psi$ and $\psi'$. Let $B_\epsilon(\psi)$ and
$B_\epsilon(\psi')$ be balls of radius $\epsilon$ centered at
$\psi$ and $\psi'$ in the ambient space, and let $T(\psi,\sigma)$ and
$T(\psi',\sigma)$ denote sets of $n\sim 2000$ tangent vectors drawn
uniformly from disks of radius $\sigma$ in $T_\psi\mathcal{M}$ and
$T_{\psi'}\mathcal{M}$. For each $v\in T(\psi,\sigma)$ we form a
proposal $\tilde{\psi} = \mathrm{Proj}_{E,A}[\psi+v]$ and count how
many such proposals land inside $B_\epsilon(\psi')$. Repeating for
$T(\psi',\sigma)$ and counting proposals that land in
$B_\epsilon(\psi)$ yields an estimate of
\begin{equation}
 \frac{\alpha(\psi\rightarrow\psi')}{\alpha(\psi'\rightarrow\psi)}
 \approx
 \frac{
  \#\{v\in T(\psi',\sigma): \mathrm{Proj}_{E,A}[\psi'+v]\in B_\varepsilon(\psi)\}
 }{
  \#\{v\in T(\psi,\sigma): \mathrm{Proj}_{E,A}[\psi+v]\in B_\varepsilon(\psi')\}
 }.
\end{equation}
The asymmetry is typically on the order of $5$–$20\%$ for
$\varepsilon\sim 0.75$, reflecting nontrivial curvature of the
high-dimensional microcanonical manifold. Including this correction is necessary to ensure convergence to the desired stationary distribution, especially when employing large step sizes $\varepsilon$.

\end{document}